\documentclass[conference]{IEEEtran}
\IEEEoverridecommandlockouts
\renewcommand\footnoterule{\kern-3pt\hrule width 0.4\columnwidth\kern 2.6pt}
\usepackage{cite}
\usepackage{amsmath,amssymb,amsfonts}
\usepackage{graphicx}
\usepackage{textcomp}
\usepackage{booktabs}
\usepackage{array}
\newcolumntype{L}[1]{>{\raggedright\arraybackslash}p{#1}}
\usepackage{xcolor}
\usepackage[hidelinks,breaklinks]{hyperref}
\usepackage{ifpdf}
\ifpdf\else\usepackage{breakurl}\fi %

\begin{document}

\title{Schemas That Hold: Verified Agentic Ingestion of Heterogeneous Scientific Tables into Relational Databases\thanks{\copyright~2026 IEEE. Personal use of this material is permitted. Permission from IEEE must be obtained for all other uses, in any current or future media, including reprinting/republishing this material for advertising or promotional purposes, creating new collective works, for resale or redistribution to servers or lists, or reuse of any copyrighted component of this work in other works. Accepted to the 6th International Workshop on Big Data \& AI-Assisted Tools, Methods, and Use Cases for Innovative Scientific Discovery (BTSD 2026), held with the 2026 IEEE International Conference on Big Data (IEEE BigData 2026), Phoenix, AZ, USA.}}

\author{\IEEEauthorblockN{Bharatha Rankothge}
\IEEEauthorblockA{Dept.\ of Mathematics\\
Colorado State University\\
Fort Collins, CO, USA\\
bharatha.rankothge@colostate.edu}
\and
\IEEEauthorblockN{Suzanne Hodgkins}
\IEEEauthorblockA{Dept.\ of Microbiology\\
The Ohio State University\\
Columbus, OH, USA\\
hodgkins.3@osu.edu}
\and
\IEEEauthorblockN{Virginia Rich}
\IEEEauthorblockA{Dept.\ of Microbiology\\
The Ohio State University\\
Columbus, OH, USA\\
rich.270@osu.edu}
\and
\IEEEauthorblockN{Michael Kirby}
\IEEEauthorblockA{Dept.\ of Mathematics\\
Colorado State University\\
Fort Collins, CO, USA\\
michael.kirby@colostate.edu}}

\maketitle

\begin{abstract}
Public scientific datasets rarely arrive analysis-ready: they come as folders of matrices, tables, and per-cohort files whose keys and encodings differ. A coding agent can load them, but the database it builds may corrupt data silently, enforce no keys or links, differ from run to run, and leave no audit trail. We present an ingestion pipeline built on one rule: LLM agents propose, deterministic code verifies, and humans decide. Agents see compact, citable evidence, never the full data; a verifier with 39 check types tests each proposal in two loops, one for structure and one for meaning; a human approves the schema and the load; and code quarantines what it cannot place and reconciles every row. On five public biomedical datasets, two held out, with five runs each, 10 of 25 first proposals failed a check and every failure was caught: the loops repaired nine and a one-line reviewer edit fixed the tenth. All 45.4 million values were loaded or quarantined with zero reconciliation gap, no planted data error passed silently, and one pipeline built differently shaped schemas without dataset-specific code. A general-purpose coding agent given the same folders preserved every value it loaded, but enforced keys and links in only 17 of 25 runs, built a different schema almost every time, stored a planted conflicting label as valid, and once reported success after loading nothing. The pipeline enforced the same keys, grains, and links in every run and recorded every decision, at similar cost (\$0.18--0.54 per run).
\end{abstract}


\section{Introduction}
Scientific research groups produce large volumes of heterogeneous data, often stored in tables, that must be cleaned, processed, and integrated before analysis. This is true whether researchers analyze their own group's data or another group's data to verify its findings. A typical public biomedical dataset holds a measurement matrix with samples in columns, a metadata table with samples in rows, and separate clinical files. Determining how to connect these tables correctly, while resolving inconsistent encodings, takes substantial time and effort.

LLM agents can now read such a folder and write and execute the code to load it into a database. Code that runs unchecked may corrupt data silently, by coercing a type, truncating a value, or dropping a row. A subtler risk is an agent that appears to succeed but quietly falls short of the task it was given. We ran a general-purpose coding agent five times on each of five public datasets and each time asked it to load every file into a normalized relational schema that preserves every value. Every run that loaded data preserved every value, yet in only 17 of 25 runs did the database enforce a key on every measurement table and require every link to its samples and features; in the rest, nothing stops a later duplicate or orphaned row. In a planted error test, when one sample carried two conflicting labels, the agent keyed the table so that both were stored as valid. And the same prompt produced a different schema almost every time. The agent alone decides when the job is done: one run exited successfully after loading nothing, and another reported keys on every table when four of its seven tables had none. Without an auditable record of what it did, these gaps surface only when someone inspects the database.

We take a different approach: \emph{agents propose, code verifies, humans decide}. Our LLM agents never write code, never see the full data, and never touch the database. Each has a narrow role, receives compact evidence measured from the files, and returns a typed proposal. A deterministic verifier checks each proposal against that evidence and, where a check needs it, the raw data, then accepts or rejects it; only accepted proposals are executed. This matters because our agents make mistakes too: ten of the 25 first proposals in our evaluation failed a check, and every failure was caught. A human reviewer approves the schema before any database access and the load before any row is inserted into the database, and rows that cannot be placed are quarantined with a reason.

\textbf{Contributions.} (1) A pipeline that separates control (a plain state machine), reasoning (narrow LLM agents), and execution (deterministic tools), with a typed conceptual schema as the contract between them (Section~\ref{sec:design}). (2) Two propose--verify loops sharing one evidence-based verifier: one checks and repairs the schema's structure, the other checks its meaning. (3) An evaluation on five public biomedical datasets, two held out from development, that tests four claims: the verifier catches the agents' errors and escalates unresolved issues to a human reviewer; every value is loaded, quarantined, or logged; one pipeline, with no dataset-specific code, handles differently shaped data; and, unlike a general-purpose coding agent given the same folders, the pipeline declares the same keys, grains, and links as database constraints on every run and keeps an audit trail of every decision, at similar cost (Section~\ref{sec:eval}).

\section{Related Work}
Repositories such as GEO store deposits as submitted, typically matrices plus metadata~\cite{geo}, so integrating them for analysis falls to each researcher. The FAIR principles call for data that are reusable with rich provenance~\cite{fair}; our pipeline supports that reuse by turning each repository deposit into a typed, keyed database with per-file provenance. To decide how tables connect, the pipeline builds on classical data profiling, measuring candidate keys and inclusion dependencies from the files~\cite{abedjan2015}. Other work automates parts of this process: Auto-Tables relationalizes non-relational tables by program synthesis~\cite{autotables}, and foundation models have been applied to entity matching and error detection~\cite{narayan2022}, schema matching~\cite{magneto}, and data standardization through agents that write and run code~\cite{cleanagent}. The closest design to ours is DataLoom~\cite{dataloom}: an LLM maps a folder of CSV files to tables, constraint discovery adds keys, a user reviews each table before PostgreSQL loads it, and the LLM suggests fixes when a load fails. We differ in three ways. Our agents' proposals must cite measured evidence and pass a deterministic verifier before a human reviewer sees them; the pipeline restructures files as needed instead of mapping each to one table, reshaping wide matrices and splitting or merging files into dimension and fact tables; and it reconciles loaded rows against the source, quarantining what it cannot place. Commercial data platforms also automate parts of ingestion: Databricks Auto Loader~\cite{autoloader} and AWS Glue~\cite{glue} infer per-file schemas, and Google's BigQuery data preparation uses an LLM to suggest cleaning steps, such as type casts and deduplication, for a user to accept~\cite{bqprep}. To our knowledge, none of these platforms infers keys and joins across a folder or verifies AI proposals against measured evidence. Finally, LLM agents built for scientific analysis~\cite{biodsa} or for running a paper's code~\cite{paper2agent} execute code the agent itself writes or chooses. Our agents never execute anything; they only emit declarative proposals that are checked deterministically, so every ingestion decision can be audited.

\section{Design}\label{sec:design}
\subsection{Principles and overview}
Our design is shaped by five rules. \emph{Data-agnostic:} no file names, column names, or domain terms are hard-coded. \emph{Agents reason; code executes:} agents emit typed specifications, and values move only through deterministic Python and PostgreSQL \texttt{COPY}. \emph{Context-bounded:} agents see compact profiles, never full matrices. \emph{Code checks, humans approve:} an agent proposes; only the verifier accepts a proposal, and only the human reviewer approves what reaches the database. \emph{Fail loud:} rows that cannot be placed are quarantined with a reason, and control flow is never an LLM decision.

\begin{figure*}[!t]
\centering
\includegraphics[width=\textwidth]{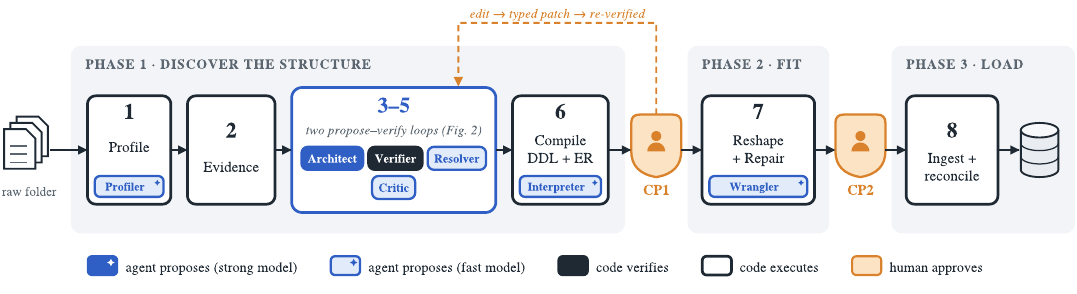}
\caption{Pipeline overview: three phases, eight stages, and two human checkpoints. The schema is approved at Checkpoint~1 (CP1) before any database access; the load is approved at Checkpoint~2 (CP2). Blue chips are LLM agents, which only propose; black and white boxes are code, which verifies or executes every proposal. Stages~3--5 are the two propose--verify loops of Fig.~\ref{fig:loops}.}
\label{fig:pipe}
\end{figure*}
The pipeline separates control, reasoning, and execution. Control is handled by a plain Python state machine that runs eight stages in three phases: \emph{discover the structure} (Stages~1--6), \emph{fit the data to it} (Stage~7), and \emph{load} (Stage~8), with a human checkpoint after each of the first two (Fig.~\ref{fig:pipe}); it persists each stage's status and can resume from any completed stage. Reasoning is done by six narrow agents (Table~\ref{tab:agents}), each a single stateless LLM call that returns a typed (Pydantic) object, not a multi-turn agent loop. In our current implementation, the Architect uses Gemini 3.1 Pro (preview); the others use Gemini 3.5 Flash-Lite. Execution is code: file readers and profilers, discovery routines for keys and joins, the verifier, the DDL compiler, the melt engine, and the loader. Agents have no tools, so they cannot read raw files or the database.

In the rest of the paper we use standard data-warehouse terms: a \emph{dimension} table describes things with an identity (samples, subjects, genes), one row each; a \emph{fact} table holds measurements, each row linking to the dimension-table rows it measures; and a fact table's \emph{grain} is the set of keys that identifies one measurement (e.g., sample and gene).

\begin{table}[!t]
\centering\footnotesize
\caption{The six agents, their proposals, and the deterministic guardrail on each.}
\label{tab:agents}
\setlength{\tabcolsep}{3pt}
\begin{tabular}{@{}L{15mm}c L{24mm}L{36mm}@{}}
\toprule
Agent & Stage & Proposal & Guardrail \\\midrule
Profiler & 1 & File shape label & One of four fixed labels; must cite headers in the card \\
Architect & 3 & Conceptual schema & 39 verifier checks \\
Resolver & 4 & Typed edit list & Kept only if fewer checks fail \\
Critic & 5 & Recodes, objections & Must match observed values; re-verified \\
Interpreter & 6 & Plain-language summary & Cannot change the DDL \\
Wrangler & 7 & Transforms (8 ops), flags & Fixed vocabulary; derived keys must exist in the parent \\
\bottomrule
\end{tabular}
\end{table}
\subsection{Evidence (Stages~1--2)}
Stage~1 builds one compact card per file: format, row and column counts, a header sample, and a summary of each column, stamped \emph{file} or \emph{sample} so agents know whether a statistic covers every row or just a sample. A file with more than 40 columns whose headers mostly share one pattern is treated as a feature matrix and summarized from a few sampled columns. The Profiler labels each file with one of four shapes (wide, long, key-value, unknown) and lists its measure columns; a proposal that gives any other label or names a column not in the card is rejected. Stage~2 measures the evidence the agents reason from and files each finding under a short label (a handle) that proposals must cite: \texttt{KEY} (candidate keys), \texttt{IND} (inclusion dependencies with coverage), \texttt{GRAIN} (parent/child grain, flagging keys unique only within a file), and others for duplicate columns, shared headers, derived columns, and whether a wide matrix should be \emph{melted} (reshaped to one row per value). The evidence contains no full matrices.

\subsection{Two propose--verify loops (Stages~3--5)}
\textbf{The Architect proposes.} It reads the bundle and returns a conceptual schema: file roles, dimension tables with keys and attributes tied to source columns, relationships with join keys, facts (grain, measures, annotations), promotions (moving a column to the parent table it describes), declared drops with reasons, and decisions it leaves open. Its operational decisions (file roles, dimension tables, joins, facts) cite the evidence handles they rest on, and the verifier rejects any claim the evidence does not support; the Architect may abstain from making a decision rather than guess.

\textbf{The verifier checks.} The proposal's declarations (partition keys, promotions, numeric extractions) are first computed from the data in code; then 39 check types in six families test the result against the evidence (Table~\ref{tab:gates}). A primary key must equal a candidate key or be a subset of one; a join needs inclusion-dependency coverage; a melt needs a wide file in the measured orientation; a promotion must be constant at the target's grain; a fact key must be unique and minimal on the data. Every metadata column must be modeled or dropped with a stated reason; a drop that loses information is flagged for the reviewer.

\begin{table}[!t]
\centering\footnotesize
\caption{Verifier check families.}
\label{tab:gates}
\setlength{\tabcolsep}{3pt}
\begin{tabular}{@{}L{14mm}L{70mm}@{}}
\toprule
Family & Example failures \\\midrule
Grounding & An attribute, measure, or annotation that names a source column that is not a profiled header \\
Keys & A key with no candidate-key support; a fact key that repeats or has a redundant column; colliding partitioned keys \\
Joins & A join without inclusion-dependency coverage; a join target that is not the parent key; mismatched key types \\
Facts & No fact for a measurement file; a melt on a non-wide file or in the wrong orientation; melted cells not matching the grid \\
Coverage & A column neither modeled nor dropped; a measurement-file column dropped although it carries information; a file assigned to no table \\
Structure & A split that contradicts \texttt{GRAIN}; a non-constant promotion; a value map that misses observed values \\
\bottomrule
\end{tabular}
\end{table}
\begin{figure}[!t]
\centering
\includegraphics[width=\columnwidth]{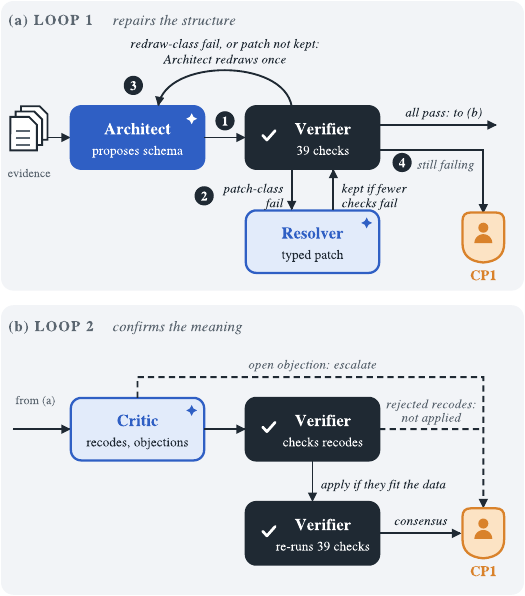}
\caption{Stages~3--5: two propose--verify loops that share one verifier. Loop~1 repairs the structure; Loop~2, which runs only on a proposal that passes every check, confirms its meaning. Numbered badges give the order of Loop~1; recodes that fail their checks are not applied and are listed for the reviewer at CP1. Dark blue: strong model; light blue: fast model; black and white: code; orange: human checkpoint.}
\label{fig:loops}
\end{figure}
\textbf{Loop~1 repairs the structure} (Fig.~\ref{fig:loops}). Patch-class failures, which are errors that a local edit can fix (grounding, coverage, joins, melt orientation, promotions, colliding local keys), go to the Resolver, which returns a patch of typed edits; code applies it to a frozen copy of the schema and re-runs the checks. The patch is kept only if fewer checks fail, unless it re-keys a table without fixing the joins to it. Redraw-class failures, which are errors that need a new schema (such as a fact grain that is not unique, or a file assigned to no table), and discarded Resolver patches get one Architect redraw guided by the failed checks; code then restores any identity column, measure, or attribute the redraw dropped without cause. Checks that still fail go to Checkpoint~1 for the reviewer to decide; nothing is forced through.

\textbf{Loop~2 confirms the meaning.} The Critic runs only on a proposal that passes every check. It judges what the checks cannot (not-applicable vs.\ missing, needs its own dimension table vs.\ is an attribute, measure vs.\ descriptive attribute, wrong grain). It then raises objections and proposes value maps, numeric extractions, or renames. A value map is applied only if it covers every observed value, and a numeric extraction is applied only if it recovers a number from at least 95\% of the observed values and that number is not the same in every row; the full verifier then runs again. Loop~2 ends at consensus: every check passes and no objection is open; otherwise the open items go to the reviewer at Checkpoint~1.

\subsection{Compile, approve, load (Stages~6--8)}
A compiler turns the verified schema into PostgreSQL DDL and an ER diagram, with a \texttt{UNIQUE} constraint on each fact's grain and \texttt{NOT NULL} fact links; the Interpreter agent writes a plain-language summary of it. At Checkpoint~1 a human reviewer approves or rejects the proposed schema, or requests edits. The reviewer's edits are held to the same standard as the agents': each request becomes a typed patch that must pass the same verifier checks before it is applied. 

Stage~7 melts wide matrices (or stages long facts as-is) and keeps missing cells as \texttt{NULL} with \texttt{value\_status=missing}. The Wrangler agent proposes transforms from a fixed vocabulary of eight operations for the issues that remain (rows of the wrong length, decisions the schema left open, keys written differently from a parent table's keys), or flags them for Checkpoint~2; a derived key is kept only if all of its values exist in the parent. Before approving the load at Checkpoint~2, the reviewer sees quarantine counts, whether each matrix's staged plus quarantined cells equal its source cells, and measurement keys with no parent row that differ from an existing parent key only in case or punctuation (possible typos). Stage~8 loads dimension tables, then streams facts through \texttt{COPY} in 5-million-row commits with keys enforced.

\subsection{Guarantees}
No LLM authors a cell value. Every source row is either reshaped into staged values or, if its length does not match the header, quarantined whole as one record that keeps its raw contents. After every commit, staged rows must equal loaded plus quarantined rows; a mismatch rolls back that commit. Each source file's SHA-256 and counts go to a provenance table. Table~\ref{tab:guar} lists how each data fault is handled.

\begin{table}[!t]
\centering\footnotesize
\caption{How the loader handles problem rows and values; none is dropped or changed silently.}
\label{tab:guar}
\setlength{\tabcolsep}{3pt}
\begin{tabular}{@{}L{30mm}L{54mm}@{}}
\toprule
Event & Response \\\midrule
Row of the wrong length & Quarantined whole (\texttt{ragged\_row}) \\
NA token or empty cell & \texttt{NULL}, \texttt{value\_status=missing} \\
Non-numeric measure & Quarantined (\texttt{unparseable\_value}) \\
Repeated fact grain & First kept; rest quarantined (\texttt{duplicate\_grain}) \\
Dimension row with no matching parent & Quarantined (\texttt{unmatched\_parent\_key}) \\
Value fails its type & \texttt{NULL} + audit row (\texttt{coerced\_to\_null}) \\
Repeated dimension key & First kept; rest quarantined (\texttt{duplicate\_pk\_conflict}) \\
Key missing from its parent table & Placeholder parent (key only) + audit row (\texttt{stub\_parent\_created}) \\
Empty fact link & Quarantined (\texttt{MISSING\_\textless{}link\textgreater{}}, e.g. \texttt{MISSING\_SAMPLE\_ID}) \\
Key differing only in case or punctuation & Flagged at Checkpoint~2 as a near-duplicate (possible typo) \\
\bottomrule
\end{tabular}
\end{table}
\section{Evaluation}\label{sec:eval}
\subsection{Setup}\label{sec:setup}
We evaluated the pipeline on five public biomedical datasets that pair omics measurements with clinical or sample metadata (Table~\ref{tab:data}). \emph{GSE22203}~\cite{gse22203} is a mouse \emph{F.~tularensis} expression series: one wide matrix (37,632 probes $\times$ 108 samples), a sample table, and a one-row series table. \emph{COVIDome}~\cite{covidome} has four long measurement tables for 105 samples (6,274,741 values: MSD cytokines, metabolomics, SOMAscan proteomics, and transcriptomics as raw counts) and an Excel sample table. \emph{GSE73072}~\cite{liu2016} has a tab-separated matrix (12,023 features $\times$ 2,886 samples), GEO metadata, and seven per-cohort clinical files for 151 subjects. Two datasets were \emph{held out}: never used during development and introduced only at the evaluation stage. One was \emph{breast.TCGA}~\cite{tcga2012}, as distributed with mixOmics~\cite{mixomics}. We used its training split: mRNA (200 genes), miRNA (184), and protein (142) blocks plus subtype labels for 150 samples (78,900 values). The other held-out dataset was HMP 16S~\cite{hmp}, also distributed with mixOmics (diverse.16S)~\cite{mixmc}. It contains 16S rRNA counts (with mixOmics' +1 offset) for 1,674 OTUs in 162 samples from 54 subjects at three body sites (271,188 values), stored with samples in rows, plus a sample table and a seven-rank taxonomy. We ran the whole pipeline, from profiling to load, five times per dataset (25 runs total). At each checkpoint we reviewed the output and, except for the escalations noted in Section~\ref{sec:claim1} and a subject table we requested in every HMP 16S run (Section~\ref{sec:claim3}), approved it unchanged.

\begin{table}[!t]
\centering\footnotesize
\caption{Datasets and run~1 of each (runs 2--5 loaded identical row counts). Tokens include thinking. A ``warn'' for a check is advisory: it is recorded and shown at Checkpoint~1 but neither blocks consensus nor triggers repair. Five-run results are in Table~\ref{tab:compare}. $^{*}$Plus three calls per Checkpoint~1 edit (Section~\ref{sec:cost}).}
\label{tab:data}
\setlength{\tabcolsep}{2pt}
\resizebox{\columnwidth}{!}{%
\begin{tabular}{@{}lrrrrr@{}}
\toprule
 & GSE22203 & COVIDome & GSE73072 & breast.TCGA & HMP 16S \\\midrule
Files (formats) & 3 (csv) & 5 (csv, xlsx) & 9 (txt, csv) & 4 (csv) & 3 (csv) \\
Measurement shape & 1 wide & 4 long & 1 wide & 3 wide & 1 wide \\
Dimension / fact tables & 3 / 1 & 6 / 4 & 4 / 1 & 4 / 3 & 4 / 1 \\
Checks pass / warn / fail & 75/2/0 & 61/1/0 & 100/2/0 & 47/0/0 & 43/2/0 \\
LLM calls (Stages~1--6 + Stage~7) & 4+1 & 7+0 & 4+1 & 4+0 & 4+0$^{*}$ \\
Tokens (k) & 71.1 & 82.5 & 72.1 & 205.2 & 54.0 \\
Fact rows loaded & 4,064,256 & 6,274,723 & 34,698,378 & 78,900 & 271,188 \\
Quarantined & 0 & 18 & 0 & 0 & 0 \\
Reconciliation gap & 0 & 0 & 0 & 0 & 0 \\
\bottomrule
\end{tabular}}
\end{table}
\subsection{Claim 1: mistakes are caught or escalated to a human reviewer}\label{sec:claim1}
Ten of the 25 first proposals failed at least one check, and every failure was caught before any table was created in the database; all five runs of each dataset converged on the same keys and fact grains (Table~\ref{tab:compare}). COVIDome's five all failed fact-key uniqueness (proteins keyed by gene symbol, which repeats across aptamers, or a metabolome key without platform), two also promoted a column that varies to a parent table, and one bound a source column twice. GSE73072's four failed on a join typed differently from the key it references (3) and on the subject key (1). One HMP 16S proposal melted the count matrix in the wrong orientation; its Resolver patch did not reduce the failures and was discarded, and an Architect redraw fixed it. Executed as proposed, each would have produced a wrong schema or a failed or lossy load; the loops repaired nine, and the tenth (COVIDome run~4) was fixed by a reviewer edit at Checkpoint~1. Every run ended with all checks passing. Across runs, the schemas differed only in table names; on COVIDome, descriptive columns such as the UniProt ID also sat in different tables.

\begin{figure}[!t]
\centering
\includegraphics[width=\columnwidth]{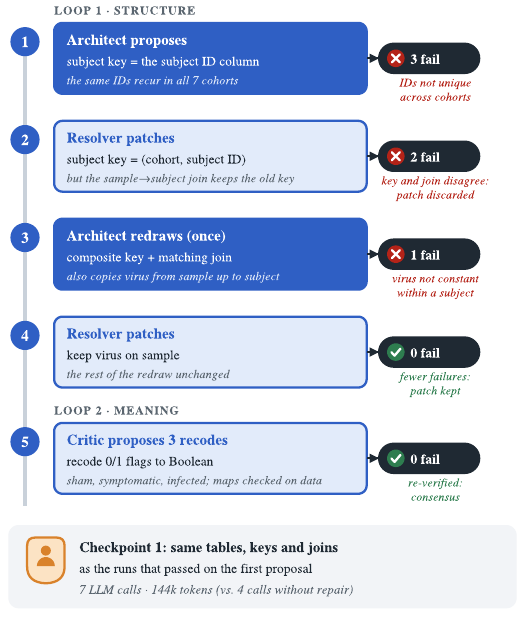}
\caption{GSE73072 run~4 through the loops of Fig.~\ref{fig:loops}; each badge is the number of failed checks. Subject IDs repeat across the seven cohort files, so a key on the ID alone is not unique. The Resolver's first patch fixed the key but not the join that references it, and was discarded. After one redraw, a second patch moved a column back from subject to sample because it varies within a subject. The Critic's recodes of three 0/1 flags to Boolean were checked against the observed values and re-verified.}
\label{fig:trace}
\end{figure}
Fig.~\ref{fig:trace} traces the GSE73072 run that proposed the subject-ID mistake: the verifier rejected the first proposal and a half-fixed patch, and one redraw plus one kept patch reached the same keys and grains as the runs that passed first time, at 7 LLM calls instead of 4. Issues the loops could not resolve were escalated to a human reviewer. On COVIDome run~4, the redraw added the platform column to all four fact keys, the minimality check failed the three where platform is constant, and with the redraw spent they went to Checkpoint~1. The reviewer's one-line feedback, ``platform only matters for the metabolite,'' became three typed \texttt{drop\_fact\_grain} operations, which were re-verified and accepted. One GSE22203 run escalated an open Critic objection to Checkpoint~1, where the reviewer approved the schema unchanged.

Across all 25 runs the reviewer's work was light. Two runs escalated items to Checkpoint~1, and the reviewer changed the schema in six, including one of those two. Every other schema and load was approved unchanged. The reviewer is expected to be a domain scientist who understands the data, not necessarily someone who writes SQL or knows the edit vocabulary: plain-language feedback suffices at Checkpoint~1, and the Architect turns it into typed edits, which must pass the same checks as its own proposals. The COVIDome feedback above is one example; in every HMP 16S run, the request ``subject\_id is the person, not the sample. Make a subject table and link samples to it.'' created the subject table; one run needed a second request, ``Move sex from sample to subject.''

\subsection{Claim 2: every value is loaded, quarantined, or logged}\label{sec:claim2}
Across all five datasets (45.4 million values in total) and in every run, staged rows equal source cells, and loaded plus quarantined rows equal staged rows exactly. Each COVIDome run quarantined the same 18 cytokine values that are Excel \texttt{\#NUM!} errors (\texttt{unparseable\_value}); no value was coerced to \texttt{NULL}. Dimension tables with no source file of their own (feature and analyte tables) were built from the distinct keys in the measurement tables, and their row counts were reported with the load, so no table was added silently. This accounting holds down to the file: every run recorded each source file's SHA-256 hash, so every loaded value can be traced to the file it came from.

\textbf{Quarantine as a backstop.} An earlier COVIDome export lacked the aptamer and Ensembl IDs, so proteins and transcripts carried only gene symbols, which repeat. The loader quarantined 32,453 rows as \texttt{duplicate\_grain} rather than merging different aptamers under one symbol, and the quarantined rows, which shared a key but had different values, showed that an identifier column was missing.

\textbf{Planted errors.} We made nine copies of the data, each with one planted error: eight in breast.TCGA and one in GSE73072, since breast.TCGA has no numeric metadata to corrupt. For each copy we re-ran only Stages~7--8, keeping the schema approved on the clean data. No error passed silently (Table~\ref{tab:planted}): five were quarantined with a reason, one was loaded as marked missing values, and three were recorded in the audit log. The case-changed ID was also flagged at Checkpoint~2. Three of the copies (a sample listed twice with different subtype labels, a sample removed from the labels, and a case-changed ID) were also run through the whole pipeline and compared with a coding agent (Section~\ref{sec:claim4}). In those runs, Stages~1--6 produced the same schema as on clean data, as it should: the planted errors corrupted individual rows and values, not the structure of the files.

\begin{table}[!t]
\centering\footnotesize
\caption{Planted errors, one per copy (CP2: Checkpoint~2).}
\label{tab:planted}
\setlength{\tabcolsep}{2pt}
\begin{tabular}{@{}L{31mm}cL{43mm}@{}}
\toprule
Planted error & Detection & Outcome \\\midrule
mRNA row one cell short & CP2 & Row quarantined, \texttt{ragged\_row} \\
\texttt{NA}, empty, \texttt{NaN} in 3 cells & load & \texttt{NULL}, \texttt{value\_status=missing} \\
\texttt{high} among numbers in the protein matrix & load & 1 value quarantined, \texttt{unparseable\_value} \\
mRNA row duplicated & load & 200 values quarantined, \texttt{duplicate\_grain} \\
Sample removed from labels & load & Placeholder sample, \texttt{stub\_parent\_created} \\
Sample listed twice with different subtypes & load & First kept, \texttt{duplicate\_pk\_conflict} \\
Sample ID blanked in a row & load & 184 values quarantined, \texttt{MISSING\_SAMPLE\_ID} \\
Sample ID \texttt{A0FJ} lowercased to \texttt{a0fj} in one file & CP2 & Near-duplicate flag; placeholder sample + audit row \\
\texttt{about 12} in numeric metadata (GSE73072) & load & \texttt{NULL}, \texttt{coerced\_to\_null} audit row \\
\bottomrule
\end{tabular}
\end{table}
\subsection{Claim 3: one fixed pipeline handles differently shaped data}\label{sec:claim3}
\begin{figure*}[!t]
\centering
\includegraphics[width=\textwidth,height=0.72\textheight,keepaspectratio]{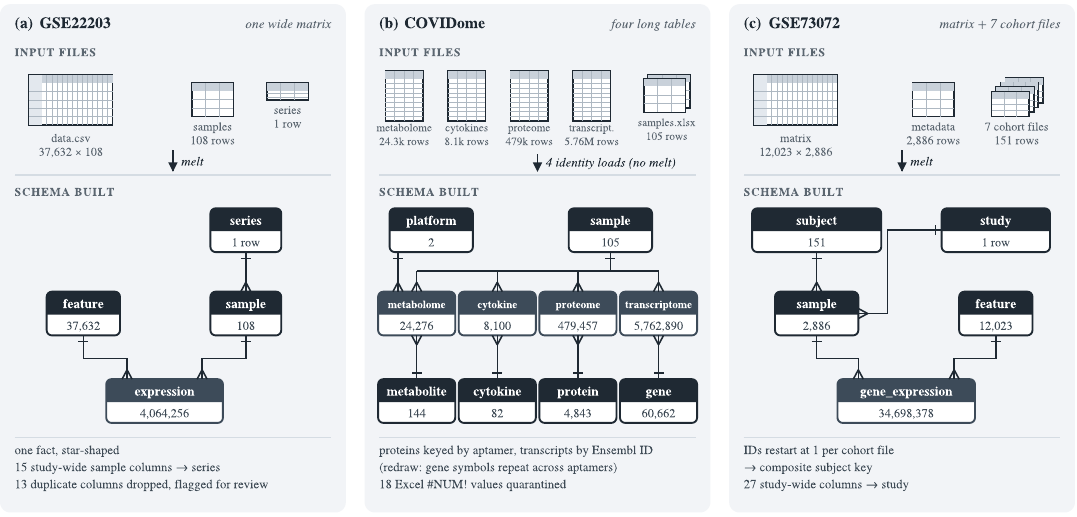}
\caption{Three datasets, three schema shapes. Top: the input files. Bottom: the schema the pipeline built, with row counts (dark headers: dimension tables; slate headers: fact tables). (a) GSE22203: one melted matrix in a star schema. (b) COVIDome: four already-long tables (6.27M values) loaded as four fact tables sharing one sample table; proteins are keyed by aptamer and transcripts by Ensembl ID; the fact tables are named with a \_fact suffix (e.g., cytokine\_fact), omitted in the figure. (c) GSE73072: a melted matrix whose subjects need a composite key.}
\label{fig:shapes}
\end{figure*}
The same pipeline produced different schema shapes without dataset-specific code (Fig.~\ref{fig:shapes}). GSE22203 became a star schema around one melted fact table, with 15 sample columns that are constant across samples promoted to the series table and 13 duplicate contact columns dropped but flagged for review. COVIDome became four fact tables sharing one sample table, loaded without a melt. GSE73072 needed the composite subject key and a one-row study table for 27 study-wide columns. The held-out breast.TCGA split, with samples in rows, became three melted fact tables with their own feature tables sharing one sample table that carries the subtype label. HMP 16S, also held out and also with samples in rows, has 1,674 OTU columns, more than PostgreSQL's 1,600-column limit; it became one melted fact table of 271,188 counts with an OTU feature table built from the taxonomy file. In each run we requested a subject table at Checkpoint~1 (54 subjects with three samples each; sex is constant per subject), a normalization choice the checks do not require; in all five runs the edit passed the checks and the table was loaded.

\subsection{Claim 4: unlike a coding agent, the pipeline declares the same keys, grains, and links on every run and records every decision}\label{sec:claim4}
To see what the pipeline adds over the model it uses, we gave a general-purpose coding agent (Gemini CLI on Gemini 3.1 Pro, the Architect's model) each of the five datasets with the same prompt: load every file into a normalized relational schema in an empty PostgreSQL database, preserve every value, and write a short report, with no internet access and nobody to answer questions. We repeated this five times per dataset, each on a fresh database, and scored every loaded database against a rubric fixed before the runs (Table~\ref{tab:compare}).

\begin{table*}[!t]
\centering\footnotesize
\caption{Five runs per dataset: the pipeline and a general-purpose coding agent given the same folders. Pipeline calls, tokens and cost cover Stages~1--7; Stage~7 made at most one Flash-Lite call per run (under \$0.001) and proposed transforms only on GSE73072: subject-key and time-point derivations that produced the same keys in every run. A run is credited with a fact key when every fact table has a primary key or unique constraint on its grain, and with required links when every link to sample and feature is NOT NULL with a foreign key. HMP 16S and COVIDome run~4 pipeline calls, tokens and cost include their Checkpoint~1 edits (three calls per edit). Tokens include thinking.}
\label{tab:compare}
\setlength{\tabcolsep}{4pt}
\begin{tabular}{@{}lrrrrr@{}}
\toprule
 & GSE22203 & COVIDome & GSE73072 & breast.TCGA & HMP 16S \\\midrule
\multicolumn{6}{@{}l}{\textbf{Pipeline}} \\
First proposal passed & 5/5 & 0/5 & 1/5 & 5/5 & 4/5 \\
Kept Resolver patch / redraw & 0 / 0 & 3 / 5 & 3 / 2 & 0 / 0 & 0 / 1 \\
Escalated to CP1 & 1 & 1 & 0 & 0 & 0 \\
Same keys and grains & 5/5 & 5/5 & 5/5 & 5/5 & 5/5 \\
Loaded, zero reconciliation gap & 5/5 & 5/5 & 5/5 & 5/5 & 5/5 \\
Rows quarantined per run & 0 & 18 & 0 & 0 & 0 \\
LLM calls, median (range) & 5 (5--5) & 7 (5--8) & 6 (5--8) & 4 (4--4) & 7 (7--11) \\
Median tokens (k) & 69 & 85 & 94 & 204 & 54 \\
Cost (\$), median (range) & 0.25 (0.19--0.27) & 0.44 (0.29--0.51) & 0.29 (0.18--0.54) & 0.41 (0.40--0.42) & 0.24 (0.19--0.33) \\\midrule
\multicolumn{6}{@{}l}{\textbf{Coding agent}} \\
Every value preserved & 5/5 & 5/5 & 5/5 & 4/5 & 5/5 \\
Fact key and required links & 3/5 & 4/5 & 2/5 & 4/5 & 4/5 \\
LLM calls, median (range) & 17 (13--26) & 25 (18--48) & 31 (21--49) & 16 (2--24) & 18 (9--28) \\
Median tokens (k) & 371 & 460 & 1,529 & 773 & 783 \\
Cost (\$), median (range) & 0.26 (0.22--0.39) & 0.39 (0.28--0.61) & 0.65 (0.53--1.26) & 0.37 (0.10--0.46) & 0.35 (0.22--0.65) \\
\bottomrule
\end{tabular}
\end{table*}
In our runs, the agent did not invent values. Every run that loaded the data preserved every source value exactly, and when it declared a key, the key was correct; it avoided the platform (COVIDome) and subject-key (GSE73072) errors that the verifier caught in the pipeline's first proposals. What it did not do reliably was make the database enforce what it got right. Only 17 of 25 runs put a key on every fact table and made every link to sample and feature NOT NULL with a foreign key; one run, for example, never linked samples to subjects. Other runs fell short in other ways: one stored every measurement as text, and one loaded nothing yet exited claiming success. The agent's schemas also changed from run to run: in five runs per dataset it produced five different layouts on COVIDome and GSE73072, four on breast.TCGA, and three on HMP 16S. In contrast, the pipeline produced identical keys and fact grains in every run of each dataset, and all 25 of its loads enforced every key and link and reconciled exactly with the source. The agent's own inspections of the data in each run show that the pipeline's evidence anticipates what an agent wants to know: of the 238 commands in which it read the source files, 235 (99\%) asked something the pipeline already answers, 174 from the evidence (19 of them on a sample), 52 in the verifier on the full data, and 9 at load.

The three planted-error copies of breast.TCGA (Section~\ref{sec:claim2}), on which we also ran the pipeline end to end, show what happens when the data itself is wrong. We ran the agent three times on each copy. On the copy in which one sample carries two subtype labels (Table~\ref{tab:planted}), all three agent runs keyed subtypes on (sample, subtype), turning the conflict into a valid many-to-many link; one run's report did not mention the conflict. The pipeline's key, derived from the evidence and approved before any row is loaded, kept the first label and quarantined the second with a reason, so the conflict was recorded rather than absorbed into the schema. On the two other copies compared (a removed label and a case-changed ID), the agent loaded the data correctly, but one run on each did not report how it resolved the error.

Across the five datasets, the agent used 4--16$\times$ the pipeline's tokens (median per dataset), but 85--95\% of its input is billed at the cached rate while the Architect's thinking is billed as output, so costs were similar: a median of \$0.26--0.65 per run for the agent and \$0.24--0.44 for the pipeline (Table~\ref{tab:compare}). The pipeline's advantage over the coding agent is thus not cost but enforced constraints, the same keys, grains, and links on every run, and an audit trail of every decision.

\subsection{Cost}\label{sec:cost}
Discovering the structure took 4 LLM calls when the first proposal passed and 5--7 when the repair loops triggered. The Wrangler added at most one. Each Checkpoint~1 edit request goes back to the Architect and through the verifier, costing at least three calls (Architect, Critic, Interpreter) and more if the edited schema fails a check and enters the repair loop; each HMP 16S edit took three. Across all 25 runs, a run used 49--205k tokens, \$0.18--0.54 at listed Gemini API prices~\cite{geminiprice}. Load time grows with the data (3~s for breast.TCGA, 14~s for HMP 16S, 2.1~min for GSE22203, 4.6~min for COVIDome, 18.6~min for GSE73072); LLM cost does not, because agents never see the data. For example, GSE73072's 34.7M values cost a median of only \$0.29 per run, while breast.TCGA's 78,900 cost \$0.41, because its 526 named feature columns enlarge the Architect prompt. In contrast, HMP 16S's 1,674 columns do not enlarge the prompt. Because the file has more than 40 columns and all 1,674 OTU columns share one pattern (OTU\_97.*), the Stage~1 profiling code treats the file as a feature matrix and summarizes it from a few sampled columns, whereas breast.TCGA's gene, miRNA, and protein names share no pattern, so every column is profiled.

\section{Discussion and Limitations}\label{sec:disc}
The checks turn plausible-but-wrong proposals into explicit failures: the mis-keyed GSE73072 schema, which would have loaded without error, failed three checks before any table was created. When the checks cannot resolve an issue, it is escalated to the human reviewer instead of being forced into the database: two runs escalated items to Checkpoint~1 (a non-minimal key and an open Critic objection), and a one-line edit from the reviewer resolved the first. The reviewer also approves every schema and every load, so nothing reaches the database purely on an agent's word. The comparison with a coding agent shows another benefit: because keys come from the evidence, are approved at Checkpoint~1, and are declared as constraints by a deterministic load, every run arrived at the same keys, grains, and links, all enforced, whereas the agent usually chose correct keys but often left them unenforced and changed its design between runs. The cost is engineering, since each check encodes a modeling rule and the evidence must measure what the check tests. Profiles of large files are computed on a sample, so a property visible only in the full file can be missed by the agents: the evidence reported no nulls in COVIDome's proteome, which has 495. Profiling every file in full with an embedded columnar engine such as DuckDB would close this gap.

\textbf{Limitations.} The coding-agent comparison uses one agent and one prompt; a different harness or a prompt that asks for constraints might enforce keys more often. Both held-out datasets loaded correctly in every run, but they rarely needed the repair loops: breast.TCGA passed on the first proposal in all five runs and HMP 16S needed one redraw, so evidence that the loops generalize beyond the development data rests on few cases. The evaluation covers five datasets, all biomedical, with one model family. Missing parents and conflicting duplicate keys are recorded at load rather than previewed at Checkpoint~2. When a fact key repeats, the loader keeps the first row in file order and quarantines the rest, so the kept value depends on file order. The verifier does not check where descriptive columns belong, which is what varied across COVIDome runs. Wide matrices with many named features enlarge the prompt and would need better summarization at larger scale. The pipeline is designed for one-time ingestion of a deposit, not as a recurring ETL process; incremental loads and schema evolution are out of scope.

\section{Conclusion}
LLM agents can ingest heterogeneous scientific tables into a relational database without being trusted to get each decision right on the first attempt, if they only propose, code verifies and executes, and humans decide what reaches the database. On public biomedical data, the pipeline caught every faulty first proposal across 25 runs and repaired all but one, which a reviewer edit fixed; reconciled 45.4 million values with the database in every run; let none of nine planted data errors pass silently; and built differently shaped schemas with no dataset-specific code. Unlike a coding agent given the same folders, it declared the same keys, grains, and links as constraints on every run and recorded every decision, at a cost of 4--11 LLM calls and \$0.18--0.54 per run. The resulting schemas hold in three senses: they hold every source value, they hold up because the database enforces their keys and links, and they hold steady from run to run. Because no part of the pipeline is specific to biomedicine, our next step is to apply it to deposits from other sciences where data arrive in the same heterogeneous folders.

\section*{Acknowledgment}
B. Rankothge was supported by NIH grant R24AI165424. AI assistants were used to write the pipeline code (Cursor), to produce Figs.~\ref{fig:pipe}--\ref{fig:shapes} from the authors' specifications (Claude), and to edit the text for clarity (Claude). The authors reviewed all AI-assisted output and take responsibility for the content.

\end{document}